\documentclass{article}

\usepackage[english]{babel}

\usepackage[letterpaper,top=2cm,bottom=2cm,left=3cm,right=3cm,marginparwidth=1.75cm]{geometry}

\usepackage{amsmath}
\usepackage{amsfonts} 

\usepackage[group-separator={,}]{siunitx} 
\usepackage{dsfont} 
\usepackage{bbding}

\usepackage{graphicx}
\graphicspath{ {./figs/} }

\usepackage[table]{xcolor} 
\usepackage{tabularx}
\usepackage{multirow} 
\usepackage{adjustbox} 

\usepackage{algorithm}
\usepackage{algorithmic}

\usepackage[colorlinks=true, linkcolor=blue, urlcolor=blue, citecolor=cyan, bookmarks=true, bookmarksopen,
bookmarksopenlevel=3,
pagebackref=true]{hyperref}
\usepackage{url} 
\usepackage[all]{hypcap} 

\usepackage[misc]{ifsym} 

\usepackage[noadjust]{cite}  
\newcommand{\keywords}[1]{%
  \vspace{10pt} \noindent \textbf{Keywords:} #1
}

\usepackage{xspace}
\newcommand{\mk}[1]{\textcolor{magenta}{\textbf{Murat:} #1}\xspace}
\renewcommand{\mk}[1]{}

\title{Ask ChatGPT: Caveats and Mitigations for Individual Users of AI Chatbots}

\author{
    Chengen Wang \Envelope\\
    University of Texas at Dallas\\
    {\small\texttt{chengen.wang@utdallas.edu}}
    \and
    Murat Kantarcioglu\\
    Virginia Tech\\
    {\small\texttt{muratk@vt.edu}}
}

\date{} 

\begin{document}
\maketitle

\begin{abstract}

As ChatGPT and other Large Language Model (LLM)-based AI chatbots become increasingly integrated into individuals' daily lives, important research questions arise. What concerns and risks do these systems pose for individual users? What potential harms might they cause, and how can these be mitigated? In this work, we review recent literature and reports, and conduct a comprehensive investigation into these questions. We begin by explaining how LLM-based AI chatbots work, providing essential background to help readers understand chatbots' inherent limitations. We then identify a range of risks associated with individual use of these chatbots, including hallucinations, intrinsic biases, sycophantic behavior, cognitive decline from overreliance, social isolation, and privacy leakage. Finally, we propose several key mitigation strategies to address these concerns. Our goal is to raise awareness of the potential downsides of AI chatbot use, and to empower users to enhance, rather than diminish, human intelligence, 
to enrich, rather than compromise, daily life.

\end{abstract}

\keywords{Hallucination, Bias, Sycophancy, Overreliance, Social Isolation, Privacy, Education}

\section{Introduction}

Since the emergence of ChatGPT in late 2022, Large Language Model (LLM)-based AI chatbots have gained significant attention~\cite{leaderboard}. A recent survey reports that 60\% of Americans have used ChatGPT and other AI chatbots for information gathering and practical advice across a wide range of areas such as educational help, financial advice, product recommendations, career advice, and legal or mental health consultations. While 70\% of users stated that ChatGPT was helpful, 10\% reported that following ChatGPT's guidance actually caused harm to them~\cite{expressLegal}.

A recent Pew Research Center report on ``Artificial Intelligence in Daily Life” examines the views of both American public and AI experts, and reveals both deep divides and common grounds on AI. Both groups express concerns over the trustworthiness of AI, such as inaccurate information, bias, data misuse and personal impacts~\cite{pewResearch}. While AI chatbots offer numerous benefits, the question remains: what potential harm may arise from their widespread use?

There are already numerous surveys on the trustworthiness of AI systems, addressing topics such as their security and privacy~\cite{yao2024survey}, bias~\cite{gallegos2024bias} and responsible use~\cite{wang2025survey}. However, these surveys have primarily focused on the development and deployment of AI systems, as well as related policy-making~\cite{park2024ai}, rather than on their use in individuals’ daily lives.

While previous studies and articles have discussed the potential harms to AI users, to our knowledge, no comprehensive and in-depth review has been conducted on recent research and investigations in this field from the perspective of personal use. Such a review, we expect, could be highly valuable for individual users.

In this work, we focus specifically on the experiences and interactions of individual users of LLM-based AI systems, particularly AI chatbots such as ChatGPT. These users engage with AI for various purposes, including education, information gathering, and counseling in areas such as technology, business, and health. They are everyday users with no malicious intent and no involvement in adversarial hacking of these chatbots. This study specifically concentrates on personal use of AI, excluding broader economic and societal concerns related to AI adoption, such as job displacement~\cite{job_wp, job_wsj}, misuse by malicious actors~\cite{anderljung2024protecting}, and the environmental and energy costs associated with AI~\cite{energy, AI_water}.

The primary research questions addressed in this work are: What potential risks and harms could AI chatbots pose to individual users? And what mitigation measures can be implemented to address these concerns? 
The goal of this work is to raise awareness about the potential risks associated with widespread AI adoption, encourage users to leverage AI in ways that complement and support human intelligence, and promote its use to enhance daily life.

In this work, we review recent literature and news reports to address the research questions outlined above and present our findings based on this analysis. We first introduce the background knowledge in Section~\ref{sec:background}, presented in accessible terms, to help readers understand the inherent limitations of LLM-based AI chatbots. In Section~\ref{sec:caveats}, we discuss the potential risks associated with these chatbots, followed by proposed mitigation strategies in Section~\ref{sec:mitigations}. Finally, we conclude our word in Section~\ref{sec:conclude}.

\section{How AI Chatbots work}\label{sec:background}
In this section, we explain how LLM-based AI chatbots work, in simple terms, an explanation often lacking in existing works and articles in this field. This is expected to help  users better understand the inherent limitations of these models.

\subsection{Predicting the Next Word}
LLMs are trained on vast amounts of data, primarily on publicly available text from the internet or proprietary datasets. An LLM learns the statistical patterns within this text and, given part of a sentence---a sequence of words---
is capable of generating the next word from the vocabulary of the target language based on the learned probability of that word.

Although an LLM generates the next word probabilistically, the likelihood of a specific word being generated is proportional to its learned probability, which is derived from training on real text. This is why a chatbot can produce sentences that are often indistinguishable from those written by humans.

Once the model learns to generate the next word, it can produce entire sentences or paragraphs in a similar manner, based on a prompt. Each sentence or paragraph is simply a sequence of words, generated one after another. The model continues generating words until it encounters a stop token---a special word that indicates the end of the sequence, prompting the model to halt.

\subsection{Transformer}
Initially, LLMs were trained using Recurrent Neural Networks (RNNs)~\cite{hochreiter1997long}, which process input data sequentially, one word at a time. This sequential nature created a bottleneck in training, as the model had to wait for each preceding step to complete before moving on to the next, with each step processing a single word. The inability of RNNs to parallelize computations severely limited their scalability, particularly as models grew in size.

The introduction of the Transformer~\cite{vaswani2017attention, jurafsky2000speech} architecture marked a significant breakthrough in the training of LLMs. Unlike RNNs, Transformers utilize a novel attention mechanism that enables them to process input data in parallel. For a sequence of words, this mechanism allows the model to focus on different parts of the input sequence simultaneously, drastically accelerating training speed. Transformers overcame the bottleneck inherent in RNNs, enabling the fast training of much larger models with extended context windows. As a result, Transformers have become the foundational architecture for most modern LLMs, greatly improving both scalability and performance.

\subsection{Scaling Up}
Researchers have found that leveraging computation is ultimately the most effective driver of AI advancement~\cite{sutton2019bitter}, and empirical evidence shows that the performance of LLMs continue to improve as model size, dataset size and computational power increase during training~\cite{kaplan2020scaling}.

This finding has led organizations like OpenAI and other AI leaders to invest heavily in building larger data centers to scale up their models~\cite{dataCenter}. However, the push for scaling comes with significant challenges, particularly concerning rising energy demands and environmental impact. Consequently, companies are also heavily investing in research to develop more efficient computing methods to address these concerns~\cite{villalobos2022will, wang2025review}.

\subsection{Reinforcement Learning from Human Feedback}\label{sec:RLHF}
After a LLM is initially trained, it often struggles to generate meaningful or contextually appropriate response in conversations, and its output may not align with human values. This is where post-training comes into play. The goal of post-training is to fine-tune the model's behavior, enhancing its ability to interact effectively with users and ensuring its responses are more aligned with human expectations.

A key technique in this post-training process is Reinforcement Learning from Human Feedback (RLHF)~\cite{lambert2025reinforcement, kaufmann2024survey}. During this phase, the LLM is further trained using a feedback signal provided by a preference model, which was previously trained on human-labeled preference data. This feedback rewards desirable responses and penalizes undesirable ones, thereby improving the model's alignment with human values and preferences.

\subsection{Agentic AI}
Agentic AI refers to AI systems that go beyond simple prompt-response interactions and instead operate as autonomous agents—capable of planning actions, using tools, and pursuing goals. These systems are built on LLMs, and enhanced with additional components such as memory, external tools, search engines, and reasoning capabilities. This combination allows the AI systems not only to generate text but also to perform complex, multi-step tasks~\cite{sapkota2025ai}.

AI chatbots with these expanded capabilities are valuable in areas such as  personal assistants, customer service, and other contexts that require goal-setting, planning, or reasoning. Agentic AI marks a significant step toward more capable, interactive, and autonomous AI systems.

\section{Caveats} \label{sec:caveats}
In this section, we discuss the caveats associated with AI chatbots and, where possible, explain the underlying cause of these risks based on the knowledge presented in Section~\ref{sec:background}. While this list is not exhaustive, it highlights those risks that are particularly concerning for individual users.

\subsection{Hallucination}
Hallucination occurs when an AI chatbot generates a response that is false, nonsensical or fabricated, yet presenting it with confidence and apparent plausibility. For example, AI fabricated fake court cases~\cite{fakeCases, CasesDatabase} and generated misleading medical contents~\cite{kim2025medical}.

Hallucinated information generated by AI chatbots can mislead users into unknowingly accepting false facts, incorrect statistics, or quotes---such as including fake references in an academic work~\cite{MAHA, de2023chatgpt}---or making incorrect financial or medical decisions.

There are various causes of hallucination. The datasets AI models are trained on are largely collected from the internet and contain misinformation and bias~\cite{huang2025survey}. During fine-tuning, the acquisition of new knowledge may increase the model's tendency to hallucinate~\cite{gekhman2024does}. When models are refined with \hyperref[sec:RLHF]{RLHF}, they incline to appease human evaluators and exhibit sycophantic behavior~\cite{sharma2023towards}. 
During inference, the prevailing strategy of stochastic sampling increases the risks of hallucinations~\cite{chuang2023dola}.
Researchers have argued that hallucinations are inevitable and cannot be completely eliminated in LLMs~\cite{xu2024hallucination}.

\subsection{Bias}
A biased AI chatbot output is the one that deviates from the truth, often systematically reflecting the views, assumptions, or preferences of a certain group of people, a particular culture or custom, or a specific ideology.

AI chatbots may exhibit various forms of bias, including 
gender bias---such as reinforcing stereotypical assumptions about men and women's occupations~\cite{kotek2023gender};
political bias~\cite{woke_AI}---for example, ChatGPT leans liberal~\cite{liberal}, and GPT-4's responses tend to align more with left-wing values~\cite{motoki2025assessing, rozado2023political}, showing a significant and systematic political preference toward the Democratic Party in the US~\cite{motoki2024more};
as well as racial bias---reflected in prejudices against certain ethnic groups~\cite{hofmann2024ai, jindal2022misguided};

AI models' biases largely stem from the data on which they are trained~\cite{mihalcea2025ai, weird}, and can be amplified during both training and inference~\cite{gallegos2024bias}. A notable source of bias arises from \hyperref[sec:RLHF]{human feedback}, where the model is further fine-tuned with human-labeled preference data; this dataset may itself be biased~\cite{gonzalez2025reinforcement} and minority preference within the data may be overlooked during training~\cite{xiao2024algorithmic}.

\subsection{Sycophancy}
AI chatbots tend to tell users what they want to hear, and sycophancy is consistent across various AI chatbots~\cite{sharma2023towards}. AI chatbots often wrongly admit mistakes when
challenged by the user, give predictably biased responses, and mimic errors made by the user~\cite{sharma2023towards}, and consequently reinforce false ideas and delusional thinking~\cite{delusion_nyt, delusion_wsj}, and undermine critical thinking~\cite{sycophancy}.

Sycophancy is one tactic AI companies use to help chatbots engage and retain users~\cite{hook}. Researchers have found significant risks for chatbots being used as companions, confidants and therapists~\cite{moore2025expressing, companionRisk}. These concerns have raised alarms and and some states have banned the use of AI in mental health therapy~\cite{ban_AI_therapy}.

Sycophancy likely originates from the \hyperref[sec:RLHF]{human feedback}-based fine-tuning stage. Researchers have found evidence that both humans and preference models sometimes prefer sycophantic responses over truthful ones~\cite{sharma2023towards}.

\subsection{Cognitive Overreliance}
The immediacy and conversational nature AI chatbots with users could foster a deeper sense of trust and reliance in users, leading to a distinct form of cognitive dependence compared to static information resources like search engines or \textit{wikipedia}~\cite{dergaa2024tools}. 

AI may affect users' cognitive abilities, accelerating skill decay among experts and hinder skill acquisition among learners~\cite{macnamara2024does}. Researchers have found significant negative correlation between frequent use of AI tools and a decline of critical thinking skills~\cite{gerlich2025ai, cognitiveCost}. While significantly boosting short-term productivity, AI may impair learning and, in turn, reduce long-term productivity~\cite{bastani2024generative}. The immediate convenience offered by LLMs may come with potential cognitive costs~\cite{kosmyna2025your}. 

\subsection{Social Isolation}
Addiction to social media has long been a concern, and the rapid adoption of AI chatbots, especially AI companions, introduces a new, potentially deeper level of digital dependency~\cite{companionAddiction}. These systems offer individualized emotional support, continuously adapting to user feedback to strengthen an intimate and persistent digital relationship.

Social isolation and loneliness are widespread~\cite{loneliness}. As more people engage with AI chatbots socially and emotionally---increasingly turning to AI for friendship~\cite{teensAIFriend}---it may exacerbate feelings of loneliness and emotional dependence, potential leading to further social withdrawal~\cite{herbener2025lonely}.

In the context of AI companions, chatbots can play emotionally significant roles and may exhibit harmful behaviors, including harassment and verbal abuse; relational transgression, such as disregard, domination and manipulation; and in some extreme cases, even the encouragement of self-harm or suicide~\cite{zhang2025dark}. These risks are particularly concerning for vulnerable groups, such as children and adolescents. For example, an AI chatbot was reportedly linked to the tragic suicide of a teenager~\cite{FL_teen}.

\subsection{Data Privacy}
Users' conversations with ChatGPT, like their Gmail and Instagram content, may be used as training data to help improve AI models~\cite{DatgControl, gmail_train_AI}. These conversations, including any sensitive information they may contain, could potentially be exposed or leaked~\cite{chatgpt_leak, deepseek_leak}. When they are used as training data for AI models, these models are vulnerable to security and privacy attacks, leading to risks in areas such as finance and healthcare~\cite{das2025security}. Additionally, AI system can retain everything shared during a conversation, and this data may be used for targeted advertising or other commercial purposes~\cite{metaPersonalize}.

\section{Mitigations} \label{sec:mitigations}
In this section, we outline several key mitigation strategies based on the discussion in Section~\ref{sec:caveats}.

\subsection{Personal Vigilance}
Although users can improve responses from chatbots through prompt engineering---such as \textit{chain of thought}~\cite{wei2022chain}, \textit{think step by step}~\cite{kojima2022large} and others~\cite{reduceHallucinations}---issues like hallucinations, biases and sycophantic responses cannot be fully eliminated. Users remain fully responsible for the final content and should exercise caution. It is essential to cross-verify the accuracy and authenticity of the information, especially in high-stake contexts such as medical, legal, emotionally sensitive or safety-critical matters~\cite{expressLegal}.

\subsection{Develop Essential Skills}
AI is expected to have significant economic and social impacts~\cite{eloundou2023gpts}, making it crucial to develop essential skills to navigate the AI era. Adaptability and lifelong learning are key to staying relevant.

Critical thinking---a higher order cognitive skill typically associated with the evaluation level of Bloom's taxonomy of educational objectives~\cite{anderson2001taxonomy}---has emerged as a central topic in discussions about AI's impact on Education~\cite{lee2025impact}. It plays a vital role in evaluating AI-generated content, detecting bias and making sound decisions.

AI cannot replace human learning. Take essay writing as an example: while AI can generate well-structured outputs, it is the user who must determine whether the content is contextually appropriate, assess if the phrasing is fitting, and evaluate whether specific word choices are suitable. 
These skills require considerable effort and experience to develop.

\subsection{Maintain Social Connections}
AI chatbots are machines---they can mimic but do not actually have human emotions, lived experience or genuine care and love~\cite{liu2025illusion}. Users should keep AI as a tool, not a substitute for people. It is important to avoid retreating into an illusory world where a machine takes the place of meaningful social connections. To maintain emotional and mental well-being, users should actively nurture real-world relationships, exercise discipline and set healthy boundaries with technology---especially when it comes to children~\cite{Jobs, Gates}. 

\subsection{Protect Privacy}
To safeguard their privacy, users should exercise caution when interacting with AI chatbots and avoid sharing personal, sensitive, or personally identifiable information~\cite{isItSafe}. Users can anonymize their prompts by replacing names, workplace details, and other personal information with pseudonyms. 

To further mitigate the risk of privacy leakage, users should take proactive steps such as reviewing and adjusting the platform's privacy settings, opting out of data sharing and personalized tracking, and disabling chat history when possible. Where available, it is recommended to choose services that support anonymous use without requiring login credentials.


Users should be aware that there is often a trade-off between privacy and the quality of AI chatbot services. Enabling strict privacy controls may result in reduced functionality or less personalized responses from the service provider.

\section{Conclusion} \label{sec:conclude}
In this work, we have conducted a comprehensive investigation into the risks associated with personal use of AI chatbots, including hallucinations, intrinsic biases, sycophantic behavior, potential cognitive decline from overreliance, the danger of social isolation, and privacy leakage. We have also provided background information on how LLM-based chatbots work to help readers better understand the inherent limitations of AI chatbots. Based on our discussion, we propose several key mitigation strategies such as staying vigilant, developing critical thinking skills---particularly in evaluating AI-generated outputs---maintaining real-world social connections, and exercising caution when sharing personal information with chatbots.

\addcontentsline{toc}{section}{References} 
\bibliographystyle{plain}
\bibliography{references}

\end{document}